\documentclass[conference]{IEEEtran}
\IEEEoverridecommandlockouts

\usepackage{cite}
\usepackage{amsmath,amssymb,amsfonts}
\usepackage{algorithmic}
\usepackage{graphicx}
\usepackage{textcomp}
\usepackage{xcolor}\usepackage{multirow}

\def\BibTeX{{\rm B\kern-.05em{\sc i\kern-.025em b}\kern-.08em
    T\kern-.1667em\lower.7ex\hbox{E}\kern-.125emX}}
\begin{document}

\title{Crypto-RV: High-Efficiency FPGA-Based RISC-V Cryptographic Co-Processor for IoT Security\\

}

\author{
	\IEEEauthorblockN{Anh Kiet Pham\textsuperscript{1}, Van Truong Vo\textsuperscript{2}, Vu Trung Duong Le\textsuperscript{1}, Tuan Hai Vu\textsuperscript{2,3}, Hoai Luan Pham\textsuperscript{1},\\ Van Tinh Nguyen\textsuperscript{4}, and Yasuhiko Nakashima\textsuperscript{1}}
	\IEEEauthorblockA{
        \textsuperscript{1}Nara Institute of Science and Technology, 8916–5 Takayama-cho, Ikoma, Nara, 630-0192 Japan.\\
        \textsuperscript{2}University of Information Technology, Ho Chi Minh City, 700000, Vietnam \\
        \textsuperscript{3}Vietnam National University, Ho Chi Minh City, 700000, Vietnam \\
        \textsuperscript{4}Le Quy Don Technical University, Ha Noi, Viet Nam.\\
		Email: pham.anh\_kiet.pf6@naist.ac.jp, take@lqdtu.edu.vn} 
}

\maketitle

\begin{abstract}
Cryptographic operations are critical for securing IoT, edge computing, and autonomous systems. However, current RISC-V platforms lack efficient hardware support for comprehensive cryptographic algorithm families and post-quantum cryptography. This paper presents Crypto-RV, a RISC-V co-processor architecture that unifies support for SHA-256, SHA-512, SM3, SHA3-256, SHAKE-128, SHAKE-256 AES-128, HARAKA-256, and HARAKA-512 within a single 64-bit datapath. Crypto-RV introduces three key architectural innovations: a high-bandwidth internal buffer (128×64-bit), cryptography-specialized execution units with four-stage pipelined datapaths, and a double-buffering mechanism with adaptive scheduling optimized for large-hash. Implemented on  Xilinx ZCU102 FPGA at 160 MHz with 0.851 W dynamic power, Crypto-RV achieves 165 times to 1,061 times speedup over baseline RISC-V cores, 5.8 times to 17.4 times better energy efficiency compared to powerful CPUs. The design occupies only 34,704 LUTs, 37,329 FFs, and 22 BRAMs demonstrating viability for high-performance, energy-efficient cryptographic processing in resource-constrained IoT environments.
\end{abstract}

\section{INTRODUCTION}
Cryptography is fundamental for ensuring confidentiality, integrity, and authenticity in modern computer systems. Hash functions such as SHA\mbox{-}256 and SHA\mbox{-}512 in the SHA\mbox{-}2 family, SHA3\mbox{-}256/512 from the SHA\mbox{-}3 standard, and the Chinese standard SM3 are widely used in digital signatures, certificate infrastructures, blockchain protocols, and integrity verification \cite{fips180}. AES\mbox{-}128 is the de facto block cipher for authenticated encryption in protocols such as TLS, VPNs, and IEEE 802.11, as well as for full-disk and file-level encryption in storage systems \cite{fips197}. In addition, the lightweight AES-based hash function HARAKA has been proposed for high-throughput, short-input hashing in advanced constructions such as hash-based signatures \cite{haraka_pqhash}. These cryptographic primitives are increasingly deployed together in complex protocol stacks for applications ranging from cloud and data-center security to automotive, industrial control, and large-scale Internet-of-Things (IoT) infrastructures.

Recent work has explored both instruction-set extensions and dedicated accelerators for symmetric cryptography on RISC\mbox{-}V platforms. \cite{nisanci_riscv_symm} systematically evaluate standardized symmetric algorithms on RISC\mbox{-}V and demonstrate that cryptography extensions achieve 1.5 times to 8.6 times speedups over software, yet their study remains at the ISA level and does not propose a concrete co-processor microarchitecture. \cite{stoffelen_riscv} optimizes AES, ChaCha20, and Keccak for RV32I through hand-crafted assembly and bit-manipulation instructions; however, all computations share a generic RISC\mbox{-}V pipeline, limiting multi-algorithm throughput and memory bandwidth. \cite{adams_riscv_gpgpu} integrate RISC\mbox{-}V cryptography extensions into a GPGPU and report up to 6.6 times speedup for AES\mbox{-}256; however, the GPU-oriented design is unsuitable for tightly coupled IoT/edge co-processors.

On the hardware accelerator side, \cite{b2ha_mwscas} propose a unified multi-hash coprocessor for SHA\mbox{-}256/BLAKE\mbox{-}256/BLAKE2s with high throughput and area efficiency through resource sharing, but it supports only three 32-bit hash functions and operates as a standalone IP without tight CPU integration. \cite{vrca_icicdt} present a versatile resource-shared cryptographic accelerator for multiple algorithms, maximizing area efficiency; however, its memory hierarchy is relatively simple and lacks optimized scheduling for large sequential hash workloads. \cite{rca_mcsoc} introduce a reconfigurable crypto accelerator with multi-core architecture and multilevel pipeline scheduling, showing substantial throughput improvements on FPGA, but at the cost of increased complexity and focus on 32-bit algorithms only. More recently, RVCP, a high-efficiency RISC\mbox{-}V co-processor, integrates high-bandwidth internal buffers and pipelined units to accelerate eight symmetric algorithms \cite{rvcp2024}, yet RVCP lacks SHA\mbox{-}3 and HARAKA support and does not employ double-buffered data scheduling optimized for large-hash or tree-hash operations. 

To overcome these limitations, this paper proposes Crypto-RV, a high-efficiency RISC\mbox{-}V co-processor architecture that accelerates a comprehensive set of cryptographic primitives: SHA\mbox{-}256, SHA\mbox{-}512, SHA3\mbox{-}256, SHA3\mbox{-}512, SM3, AES\mbox{-}128, and HARAKA, within a unified, tightly coupled design. Crypto-RV introduces three key architectural ideas: a high-bandwidth internal buffer organization (128\,$\times$\,64-bit) that minimizes memory traffic and sustains throughput for iterative hash computations; cryptography-specialized execution units with carefully balanced four-stage pipelines that share functional resources across multiple hash and block-cipher algorithms while preserving a short critical path; and a double-buffering mechanism with adaptive data scheduling tailored for large-hash operations, significantly reducing latency for long messages and tree-based constructions. Implemented on an FPGA SoC, Crypto-RV achieves substantial improvements in latency, throughput, and energy efficiency compared with baseline RISC\mbox{-}V cores, powerful CPU implementations, and prior accelerator designs, while also providing a flexible hardware platform that can be extended in future work to support hash-based post-quantum schemes such as SPHINCS+.

The remainder of this paper is organized as follows: Section II presents the Crypto-RV architecture. Section III shows the experimental results, and Section IV concludes the paper.
\section{PROPOSED CRYPTOGRAPHY RISC-V CO-PROCESSOR}
\begin{figure}
\centering
\includegraphics[width=0.5\textwidth]{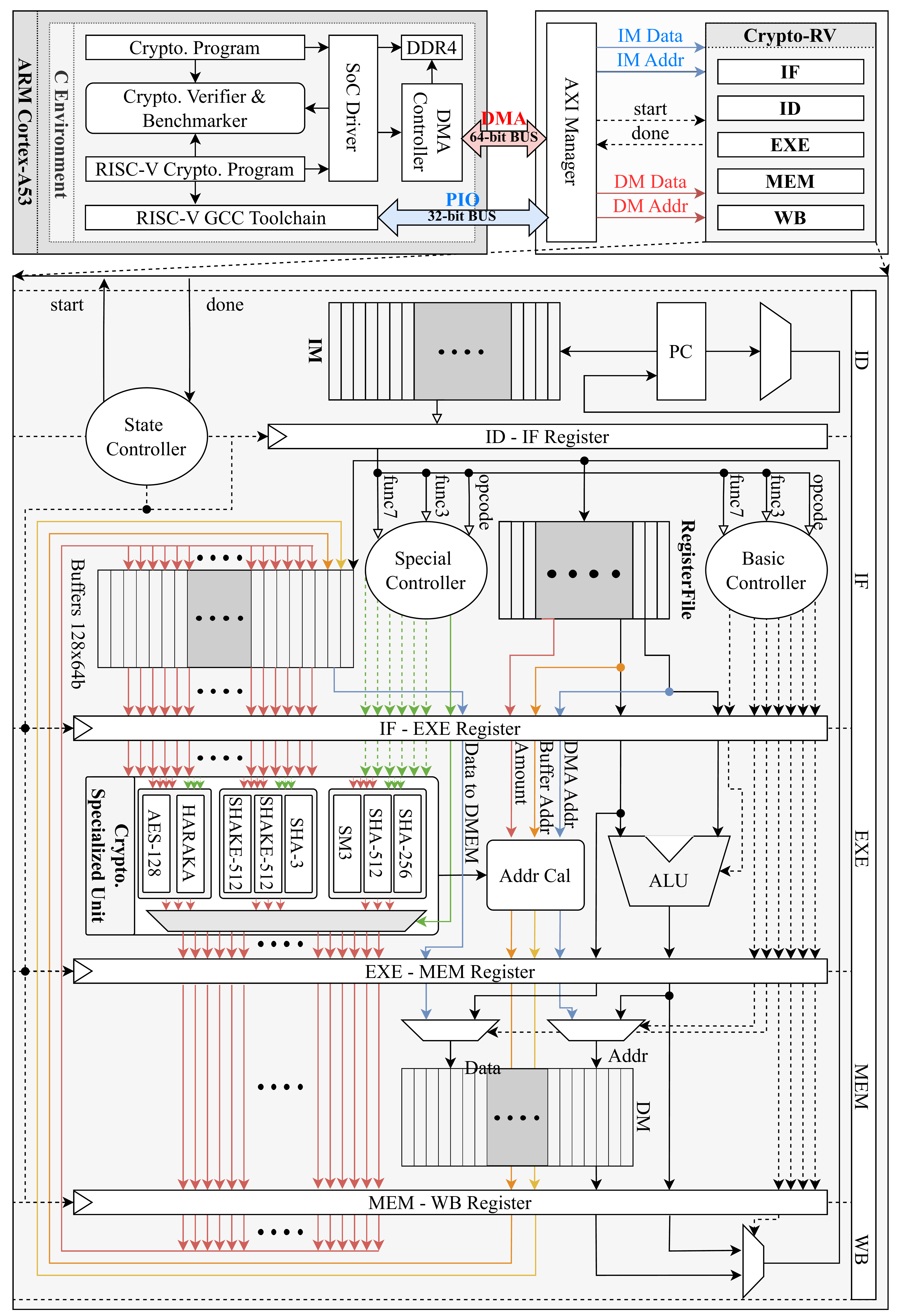}
\caption{Overview Crypto-RV Architecture on ZCU102 FPGA SoC} 
\vspace{-5mm}
\label{fig1}
\end{figure}
\begin{figure*}
\centering
\includegraphics[width=0.9\textwidth]{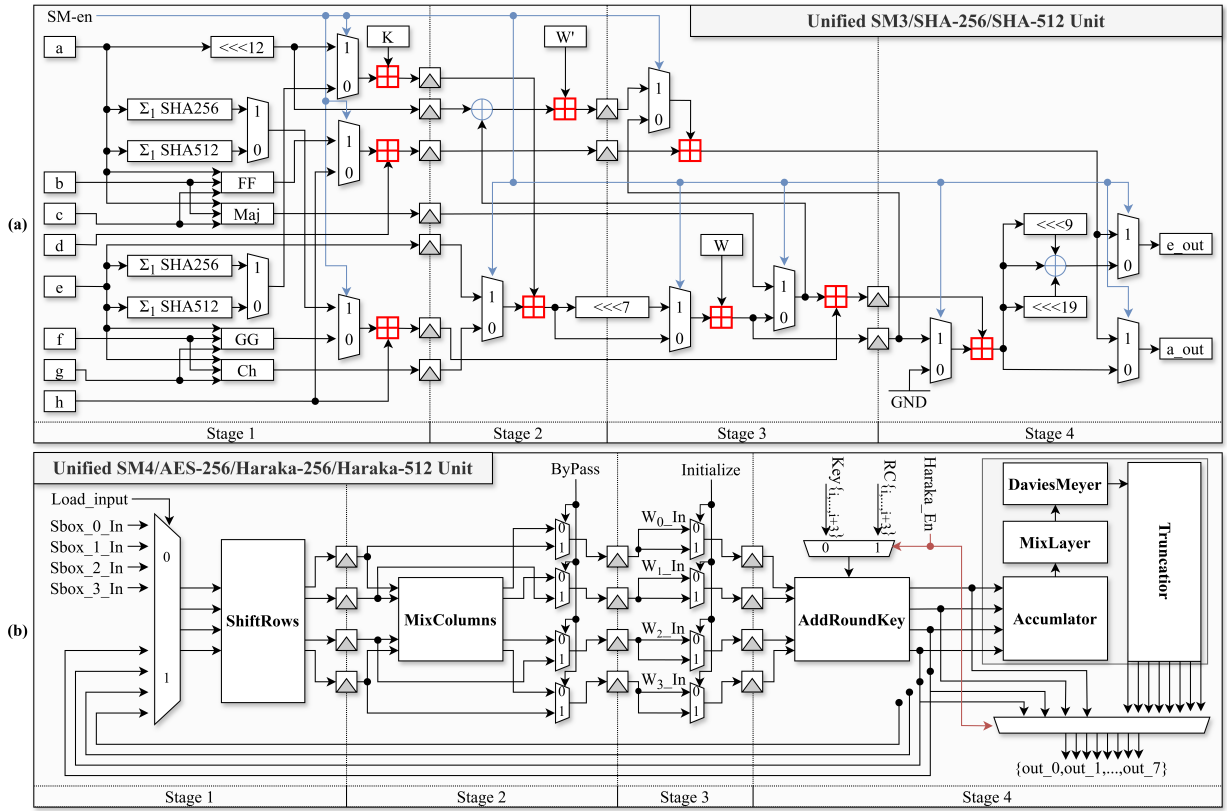}
\caption{Propose (a) Unified SM3/SHA-256/SHA-512 Unit, (b) Unified AES-128/Haraka-256/Haraka-512 Unit} 
\vspace{-5mm}
\label{fig2}
\end{figure*}
\subsection{System Architecture and Implementation Platform}
Fig. \ref{fig1} shows the system-level integration of Crypto-RV on Xilinx ZCU102 FPGA. The platform comprises two parts: the Processing System (PS) with ARM Cortex-A53 CPU running Linux, and the Programmable Logic (PL) hosting Crypto-RV. The PS executes three software components: a golden-reference crypto program for functional verification, a RISC-V crypto program that offloads kernels to Crypto-RV, and a benchmarking framework. The RISC-V GCC toolchain compiles programs into instructions while a SoC driver manages data transfers.

An AXI Manager bridges PS and Crypto-RV through two interfaces: a 64-bit DMA channel transfers bulk data between DDR4 and on-chip data memory (DM), while a 32-bit PIO interface delivers configuration and control signals to instruction memory (IM) and control registers. Once initialized, Crypto-RV's five-stage RISC-V pipeline (IF, ID, EXE, MEM, WB) autonomously fetches instructions and accesses data without further software intervention. Within the core, a state controller and custom instruction decoder manage the 128×64-bit internal buffer array and crypto-specialized unit. The buffer serves as high-bandwidth storage for constants and intermediate values, while the specialized unit implements pipelined engines for SHA-256/512, SM3, SHA3-256, SHAKE-128/256, AES-128, and HARAKA-256/512 via custom instructions. An address-calculation block manages data movement between DM, buffers, and the specialized unit, enabling full exploitation of internal bandwidth while maintaining RISC-V compatibility.
\subsection{Internal buffer for High performance}
In conventional RISC-V cores, cryptographic kernels repeatedly load intermediate states and message blocks from memory into registers, perform arithmetic operations, and spill results back, a pattern that dominates execution for hash functions processing tens of rounds per block. This creates severe memory bottlenecks: 70-85\% of cycles are spent on load/store operations rather than cryptographic computation, a problem intensified for multi-hash workloads (Merkle trees, SPHINCS$^+$) where intermediate values shuttle repeatedly between core and memory.

Crypto-RV addresses this through a dedicated 128×64-bit internal buffer tightly coupled to the execution pipeline. Rather than repeatedly accessing memory, the processor initializes buffers once with message words and constants, maintaining all intermediate states on-chip throughout the round sequence. Custom data-movement instructions enable bulk transfers of up to 128 words in single operations, decoupling high-bandwidth intra-round data reuse from low-bandwidth off-chip accesses. This architecture drastically reduces load/store instruction count, enables crypto-specialized units to one pipeline iteration per cycle after warm-up, and shares buffer layout across all algorithms for seamless data reuse without returning to external memory, delivering 17.42×–58.15× latency reduction compared to baseline RISC-V.
\subsection{Cryptography Specialized Unit}
The Cryptography Specialized Unit in Crypto-RV consists of three unified engines that execute multiple algorithms within deeply pipelined datapaths. The SM3/SHA-256/SHA-512 and AES-128/Haraka units are implemented as four-stage pipelines, while the SHA3-256/SHAKE-256/SHAKE-512 engine uses a two-stage unrolled structure to balance latency and critical path for sponge-based permutations.
\subsubsection{Unified SM3/SHA-256/SHA-512 Unit}
SM3 and SHA-2 (SHA-256, SHA-512) are Merkle–Damgard hash functions with similar iterative structures but differing word sizes (32-bit vs 64-bit), round counts (64/80), and Boolean operations. Related hardware implementation requires separate datapaths for each algorithm, resulting in significant area overhead and resource underutilization. 

Crypto-RV addresses this with a unified SM3/SHA-256/SHA-512 engine sharing functional units across all three algorithms. The design comprises a Message Expander (ME) that expands 16 input words into 64 or 80 round words, a Message Compressor (MC) performing core compression with shared adders and mode-select multiplexers, and a Value Rotator (VR) storing results back to buffers. Organized as a four-stage pipeline with one adder per stage, this balanced partitioning maintains a critical path matching the baseline RISC-V ALU. In SHA-512 mode, the unit processes one 1024-bit block per cycle; in SHA-256/SM3 mode, the 32-bit datapath processes two blocks in parallel, effectively doubling throughput while sharing over 80\% of arithmetic and logic resources.
\subsubsection{Unified AES-128/Haraka-256/Haraka-512 Unit}
AES-128 and the Haraka hash family (Haraka-256/512) are both AES-based primitives that rely on substitution-permutation networks (SPN) for diffusion and confusion. AES-128 operates on 128-bit blocks through 10 rounds of SubBytes, ShiftRows, MixColumns, and AddRoundKey transformations, while Haraka processes 32/64-byte inputs through multiple AES rounds in a sponge-like structure. A critical challenge for Haraka acceleration is that real-world deployments require full functionality: computing round constants (RC) from seed key (SK) and public key (PK) before hashing input data. Without RC acceleration, Haraka implementations can only speed up 30\% of total computation, rendering hardware acceleration ineffective for practical SPHINCS+ signatures or similar schemes.

Crypto-RV solves this bottleneck with a unified AES-128/Haraka-256/Haraka-512 engine featuring a four-stage pipeline that accelerates all Haraka computations, as shown in Fig. \ref{fig2}(b). The pipeline handles SubBytes (Stage 1), ShiftRows/MixColumns (Stage 2), AddRoundKey (Stage 3), and output accumulation (Stage 4) for both AES encryption/decryption and full Haraka sponge operations including RC generation from SK/PK. Mode-select multiplexers enable seamless switching between AES 10-round block processing and Haraka's variable round counts (32 for Haraka-256, 64 for Haraka-512), with dedicated control logic for RC precomputation. This comprehensive acceleration eliminates Haraka's RC bottleneck, delivering 3$\times$ higher effective throughput than partial implementations while sharing over 75\% of resources across all three algorithms, making Crypto-RV uniquely suited for complete hash-based cryptographic workloads.
\subsubsection{Unified SHA3-256/SHAKE-128/SHAKE-256 Unit}
\begin{figure}
\centering
\includegraphics[width=0.5\textwidth]{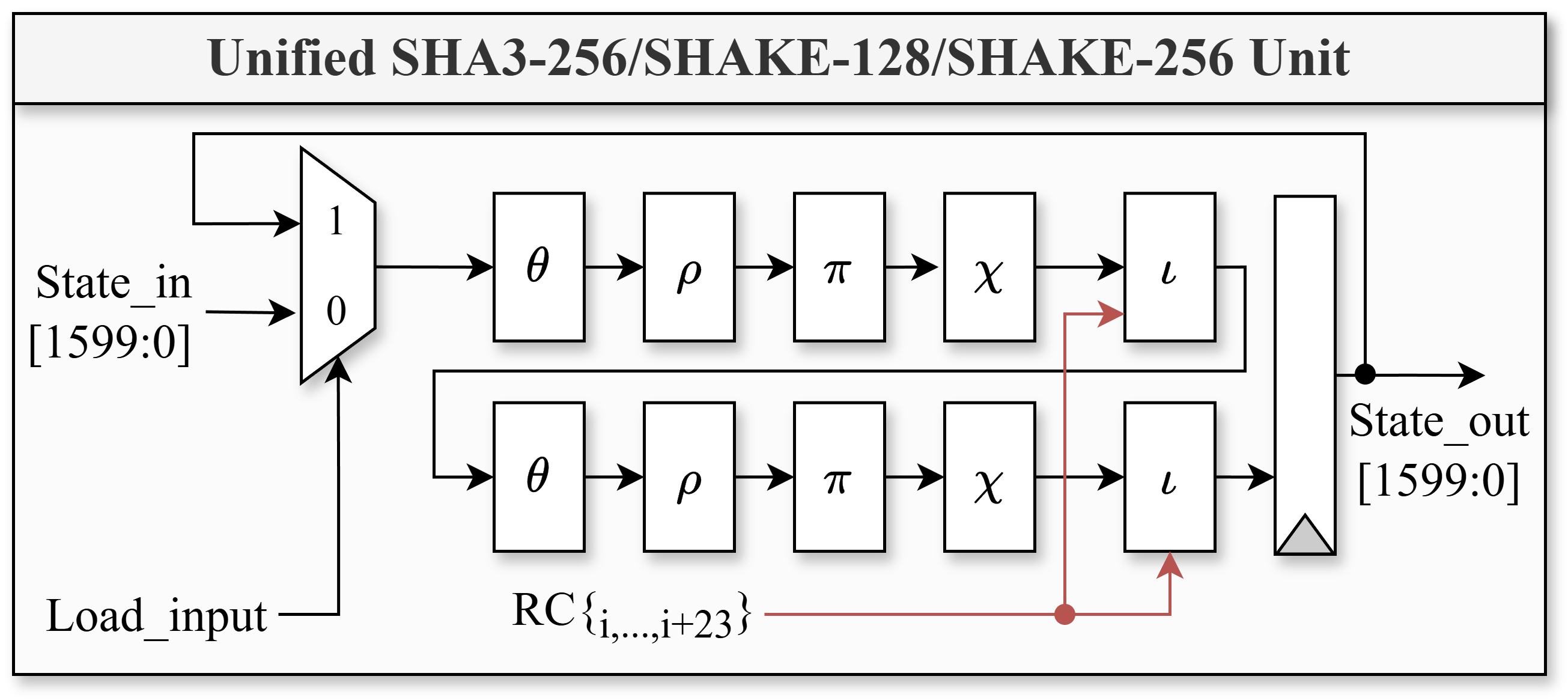}
\caption{Unified SHA3-256/SHAKE-128/SHAKE-256 Unit.} 
\vspace{-5mm}
\label{fig_sha3}
\end{figure}
\begin{figure*}[t]
\centering
\includegraphics[width=1\textwidth]{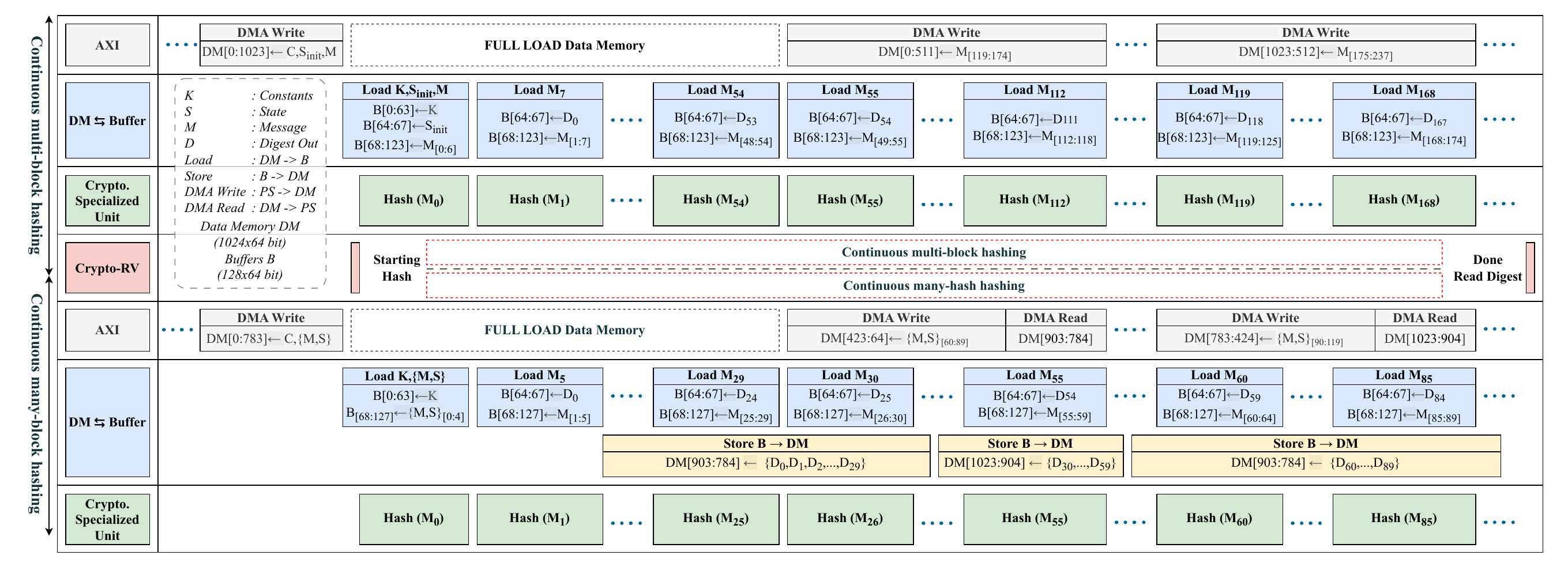}
\caption{Double-buffering schedule.} 
\vspace{-5mm}
\label{double_buffering}
\end{figure*}
SHA3-256 and the SHAKE extendable-output functions rely on a 1600-bit sponge permutation requiring 24 sequential rounds of $\theta$, $\rho$, $\pi$, $\chi$, and $\iota$ operations. A direct hardware mapping of the 24-round permutation is challenging due to the large state size and complex round function, which creates a long critical path and high latency even with pipelining.

Crypto-RV addresses this through a unified SHA3/SHAKE engine that selectively unrolls two consecutive Keccak rounds per clock cycle within a carefully optimized combinational datapath, reducing effective iteration depth from 24 to 12 rounds while preserving timing closure. As shown in Fig. \ref{fig_sha3}, the design stacks two parallel round pipelines ($\theta \to \rho \to \pi \to \chi \to \iota$ stages), each processing half the permutation while sharing round constants $\text{RC}_{i,...,i+23}$. This two-round unroll halves latency compared to a single-round design while maintaining a critical path aligned with the baseline RISC-V ALU, enabling 160~MHz operation. Mode-select multiplexers seamlessly switch between SHA3-256 fixed-length output and SHAKE-128/256 variable-length squeeze phases, all within the same hardware. The design shares state registers and constant generation across all three algorithms, delivering high performance for all Keccak-based functions while preserving timing closure.

\subsection{Double-Buffering for Continuous Big-Hash Processing}
Modern hash accelerators struggle to maintain continuous computation when processing large workloads due to a fundamental throughput mismatch between specialized crypto cores and memory subsystem bandwidth. While hash computations proceed at high throughput, data movement via DMA becomes a significant bottleneck, causing the majority of execution cycles to be consumed by memory transfers rather than cryptographic operations. This memory-bound behavior severely restricts core utilization and prevents sustained pipeline operation.

Crypto-RV overcomes this through hierarchical double-buffering between a 1024×64-bit Data Memory (DM) and 128×64-bit internal Buffer (B). DM preloads the entire workload (constants K, initial states S$_0^i$, messages M$_i$) at startup; fresh data streams into the buffer while computation proceeds, achieving perfect compute-DMA overlap with $T_{total} \approx T_{compute}$. The design supports two modes: long-message chaining maintains constants and state on-chip while message blocks stream via sliding window (S$_n$ = Hash(S$_{n-1}$ + M$_n$)), while many-hash workloads treat DM as a circular buffer processing 8 instances per batch, immediately streaming digests to output. This continuous operation achieves efficient core utilization, significantly reducing execution time for hash-intensive PQC workloads such as SPHINCS+ signature generation compared to software implementations, while the architecture generalizes across diverse 
cryptographic algorithms.
\section{VERIFICATION AND EVALUATION RESULTS}
\begin{figure}[t]
\centering
\includegraphics[width=0.5\textwidth]{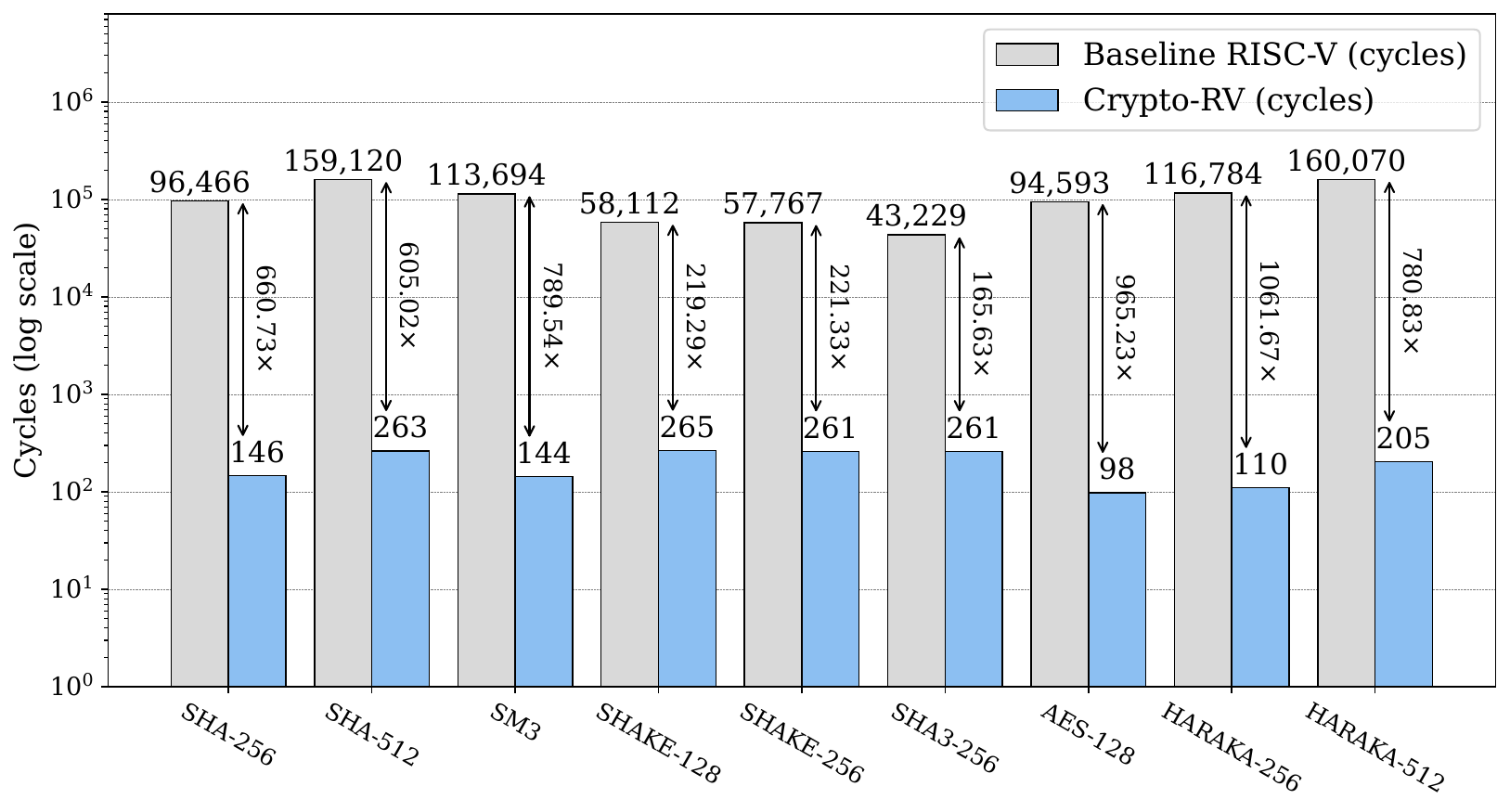}
\vspace{-5mm}
\caption{Total cycles per algorithm: Crypto-RV vs RISC-V baseline.} 
\vspace{-5mm}
\label{cycles}
\end{figure}
\subsection{Verification and Implementation Results on FPGA}
To validate Crypto-RV functionality, we implemented the complete SoC on Xilinx ZCU102 FPGA, successfully processing 100\% of 10,000,000 test cases across SHA-256/512, SM3, SHAKE-128/256, SHA3-256, AES-128, HARAKA-256/512 at \textbf{160 MHz}. The design occupies \textbf{34,704} LUTs, \textbf{37,329} FFs, and \textbf{22} BRAMs with total SoC power consumption of \textbf{4.03 W} (\textbf{3.33 W} dynamic, \textbf{0.7 W} static), of which Crypto-RV contributes \textbf{0.851 W} dynamic power, show in Table \ref{utilization}.

The cycle count comparison in Fig. \ref{cycles} demonstrates substantial speedups over baseline RISC-V: SHA-256/512/SM3 achieve 660×/604×/789× improvements respectively, SHAKE-128/256 reach 220× speedup, and AES-128/HARAKA-256/512 achieve the most dramatic gains at 965×/1061×/780× respectively. These results confirm that the double-buffering architecture and pipelined design achieve significant cycle reduction while maintaining minimal area and power overhead, validating Crypto-RV's efficiency for cryptographic acceleration.
\begin{table}[t]
\caption{Detail utilization and power consumption of Crypto-RV specialized units}
\label{utilization}
\begin{center}
\renewcommand{\arraystretch}{1.3}
\begin{tabular}{|c|c|c|c|}
\hline
\multirow{2}{*}{\textbf{Algorithms}}
& \multicolumn{2}{|c|}{\textbf{Hardware Resource}}
& \multirow{2}{*}{\textbf{Power (W)}} \\
\cline{2-3}
& \textbf{LUT} & \textbf{FF} & \\
\hline

SHA2-256   & \multirow{3}{*}{3,666} & \multirow{3}{*}{2,096} & \multirow{3}{*}{0.127} \\
\cline{1-1}
SHA2-512   &                        &                        &                        \\
\cline{1-1}
SM3        &                        &                        &                        \\
\hline

SHA3-256   & \multirow{3}{*}{5,329} & \multirow{3}{*}{3,724} & \multirow{3}{*}{0.200} \\
\cline{1-1}
SHAKE-128 &                        &                        &                        \\
\cline{1-1}
SHAKE-256   &                        &                        &                        \\
\hline

AES-128        & \multirow{3}{*}{11,308} & \multirow{3}{*}{10,895} & \multirow{3}{*}{0.491} \\
\cline{1-1}
HARAKA-256     &                        &                        &                        \\
\cline{1-1}
HARAKA-512     &                        &                        &                        \\
\hline
\end{tabular}
\end{center}
\vspace{-5mm}
\end{table}
\subsection{Comparison with state-of-the-art CPUs}
To evaluate Crypto-RV's efficiency, we compare it with Intel i9-10940X (31.5 W), Intel i7-12700H (23.6 W), and ARM Cortex-A53 (2.7 W). Fig.~\ref{power_efficiency} shows energy efficiency results. Crypto-RV achieves power efficiency from 62.76 to 187.08 Mbps/W across all algorithms.

\textbf{Intel i9-10940X:} Crypto-RV provides \textbf{4.0 times to 11.8 times} better efficiency. Notably, SHA-512 reaches 11.8 times (187.08 vs. 15.89 Mbps/W), SHAKE-128 achieves 11.4 times (145.05 vs. 12.74 Mbps/W), and SHAKE-256 provides 11.6 times (147.27 vs. 12.67 Mbps/W). Hash functions (SHA-256, SM3) deliver 8.8 times to 10.2 times improvements. Post-quantum primitives show 4.8 times to 7.3 times gains, with HARAKA-256 and HARAKA-512 achieving 4.8 times and 5.8 times respectively.

\textbf{Intel i7-12700H:} Crypto-RV delivers \textbf{3.2 times to 9.5 times} improvements. SHA-512 reaches 9.5 times (187.08 vs. 19.74 Mbps/W), SHAKE variants achieve 9.2 times to 9.3 times, and SM3 provides 8.1 times. AES-128 shows 3.2 times efficiency gain, while HARAKA variants deliver 7.1 times to 7.6 times improvements.

\textbf{ARM Cortex-A53:} Crypto-RV achieves \textbf{1.2 times to 3.2 times} efficiency gains. SHA-512 reaches 3.2 times (187.08 vs. 59.24 Mbps/W), SHAKE-128/256 provide 3.1 times, and SM3 delivers 2.6 times. SHA-256 shows modest 1.2 times gain due to ARM A53's high efficiency on this algorithm. HARAKA variants achieve 2.5 times to 3.2 times improvements.

These results validate specialized hardware acceleration effectiveness for post-quantum (HARAKA) and memory-intensive algorithms (SHAKE), making Crypto-RV ideal for power-constrained IoT systems.
\subsection{Comparison with related RISC-V work}
Table \ref{cycle_compare_fpga} presents a comprehensive execution cycle comparison between Crypto-RV and related RISC-V cryptographic accelerators, highlighting the superior throughput of Crypto-RV. Crypto-RV demonstrates a significant reduction in cycles per byte across various cryptographic algorithms, outperforming prior designs. The detailed results are as follows:
\begin{figure}
\centering
\includegraphics[width=0.5\textwidth]{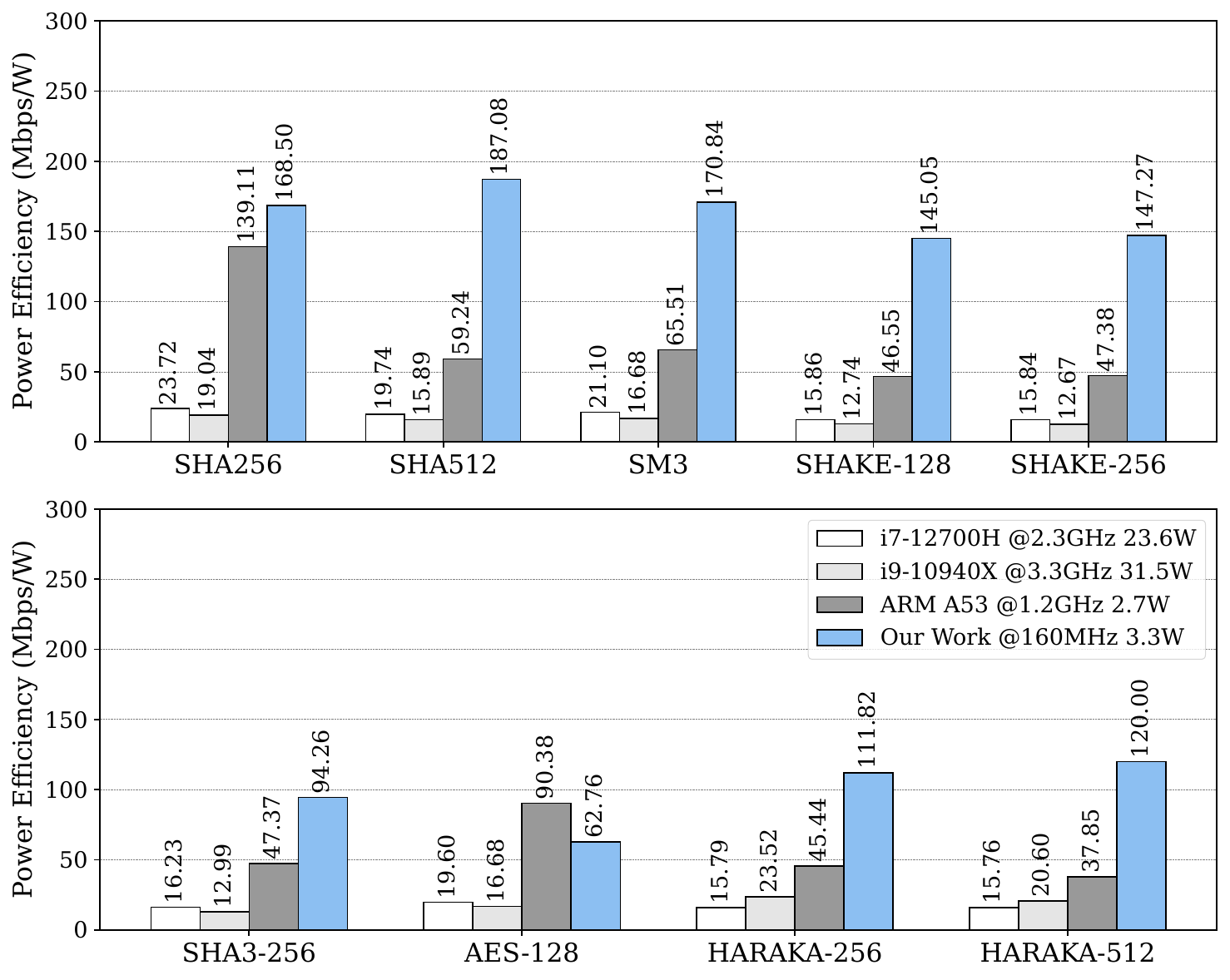}
\caption{Power efficiency comparison between Crypto-RV and powerful CPUs.} 
\vspace{-5mm}
\label{power_efficiency}
\end{figure}
\begin{itemize}
\item \textbf{SHA-256:} Crypto-RV achieves a throughput that is \textbf{56.70 times} higher than the reference design \cite{10.1145/3546000.3546013}, with 146 cycles (2.28 cycles/byte).

\item \textbf{SHA-512:} Crypto-RV shows an improvement ranging from \textbf{25.54 times to 1,291.68 times} compared to reference designs \cite{nisanci2022symmetric,10.1007/978-3-030-12942-2_21}, executing in 263 cycles (2.05 cycles/byte).

\item \textbf{SM3:} Crypto-RV is \textbf{49.18 times} faster than the reference design \cite{10102649}, completing in 144 cycles (2.25 cycles/byte).

\item \textbf{SHAKE-128/256:} Crypto-RV achieves 265 and 261 cycles (2.61--2.65 cycles/byte), delivering consistent efficiency across sponge-based functions.

\item \textbf{SHA3-256:} Crypto-RV completes in 261 cycles (4.08 cycles/byte), enabling efficient post-quantum cryptography support.

\item \textbf{AES-128:} Crypto-RV demonstrates superior performance ranging from \textbf{14.23 times to 391.10 times} better than reference designs \cite{cryptoeprint:2020/1123,AES2}, executing in 98 cycles (6.13 cycles/byte).

\item \textbf{HARAKA-256/512:} Crypto-RV achieves 110 and 205 cycles (3.44 and 3.20 cycles/byte), providing critical acceleration for hash-based post-quantum signatures.
\end{itemize}

Overall, Crypto-RV's consistent 2.0 to 4.1 cycles/byte across all algorithms underscores the effectiveness of the unified architecture in transforming memory-bound operations into compute-bound execution. The comparison results demonstrate Crypto-RV's significantly enhanced throughput compared to existing RISC-V designs, particularly for hash functions (SHA-256, SHA-512, SM3) and post-quantum primitives (HARAKA), enabling sustainable high-throughput processing without memory bottlenecks.
\begin{table}[t]
\caption{Execution cycle comparison between\\prior works and Crypto-RV}
\label{cycle_compare_fpga}
\centering
\renewcommand{\arraystretch}{1}
\setlength{\tabcolsep}{6pt}
\begin{tabular}{|c|c|c|c|c|}
\hline
\textbf{Reference} & \textbf{Algorithm} & \textbf{Cycles} & \textbf{Cycles/Byte} & \textbf{Improve} \\
\hline

{\cite{10.1145/3546000.3546013}}      & SHA-256   & 8,278   & 129.30 & 56.70$\times$ \\
\hline
{\cite{nisanci2022symmetric}}         & SHA-512   & 13,975  & 109.20 & 53.14$\times$ \\
\hline
{\cite{10.1007/978-3-030-12942-2_21}} & SHA-512   & 339,712 & 2.65   & 1,291$\times$ \\
\hline
{\cite{10102649}}                     & SM3       & 7,082   & 110.70 & 49.18$\times$ \\
\hline
{\cite{cryptoeprint:2020/1123}}       & AES-128   & 10,306  & 644.10 & 105.16$\times$ \\
\hline
{\cite{AES2}}                         & AES-128   & 38,328  & 2.40   & 391.10$\times$ \\
\hline
\multirow{4}{*}{\cite{10908294}}      & SHA-256   & 2,495   & 38.98  & 17.09$\times$ \\
\cline{2-5}
                                     & SHA-512   & 6,716   & 52.47  & 25.54$\times$ \\
\cline{2-5}
                                     & SM3       & 2,038   & 31.84  & 14.15$\times$ \\
\cline{2-5}
                                     & AES-128   & 5,590   & 349.38 & 57.04$\times$ \\
\hline
{\cite{VanTinhNguyen202522.20250329}} & AES-128   & 1,395   & 87.19  & 14.23$\times$ \\
\hline

\multirow{9}{*}{\textbf{Crypto-RV}}   & \textbf{SHA-256}    & \textbf{146} & \textbf{2.28} & \textbf{-} \\
\cline{2-5}
                                     & \textbf{SHA-512}    & \textbf{263} & \textbf{2.05} & \textbf{-} \\
\cline{2-5}
                                     & \textbf{SM3}        & \textbf{144} & \textbf{2.25} & \textbf{-} \\
\cline{2-5}
                                     & \textbf{SHAKE-128}  & \textbf{265} & \textbf{2.65} & \textbf{-} \\
\cline{2-5}
                                     & \textbf{SHAKE-256}  & \textbf{261} & \textbf{2.61} & \textbf{-} \\
\cline{2-5}
                                     & \textbf{SHA3-256}   & \textbf{261} & \textbf{4.08} & \textbf{-} \\
\cline{2-5}
                                     & \textbf{AES-128}    & \textbf{98}  & \textbf{6.13} & \textbf{-} \\
\cline{2-5}
                                     & \textbf{HARAKA-256} & \textbf{110} & \textbf{3.44} & \textbf{-} \\
\cline{2-5}
                                     & \textbf{HARAKA-512} & \textbf{205} & \textbf{3.20} & \textbf{-} \\
\hline

\end{tabular}
\vspace{-5mm}
\end{table}
\section{CONCLUSION}
This paper presents Crypto-RV, a high-efficiency RISC-V co-processor that unifies SHA-2, SHA-3, SM3, AES-128, and HARAKA within a single 64-bit datapath. Crypto-RV achieves multi-algorithm flexibility with minimal area overhead, sustains high-throughput computation through hierarchical buffering, and eliminates memory stalls via continuous double-buffering, transforming cryptographic operations from memory-bound to compute-bound execution. The architecture demonstrates exceptional efficiency for IoT and edge computing systems where power and area constraints are critical. Future work will extend Crypto-RV to fully accelerate SPHINCS+ signature generation through specialized tree-hash cores and optimized Merkle-layer state management, establishing it as a unified platform for quantum-safe IoT security.
\section*{ACKNOWLEDGMENT}
This research was funded by JST-ALCA-Next (JPMJAN23F4) and the Vietnam National Foundation for Science and Technology Development (NAFOSTED) under Grant 102.01-2025.50.
\bibliographystyle{IEEEtran}
\bibliography{references.bib}

@misc{fips180,
  author       = {{NIST}},
  title        = {{FIPS} 180-4: Secure Hash Standard (SHS)},
  howpublished = {Federal Information Processing Standards Publication},
  year         = {2015}
}

@misc{fips197,
  author       = {{NIST}},
  title        = {{FIPS} 197: Advanced Encryption Standard (AES)},
  howpublished = {Federal Information Processing Standards Publication},
  year         = {2001}
}

@inproceedings{haraka_pqhash,
  author    = {Stefan K{\"o}lbl and Martin Lauridsen and Christian Rechberger and Peter Schwabe and Gregor Seiler},
  title     = {Haraka: Efficient Short-Input Hashing for Post-Quantum Applications},
  booktitle = {Progress in Cryptology -- {ASIACRYPT} 2016},
  series    = {LNCS},
  volume    = {10031},
  pages     = {353--377},
  year      = {2016},
  publisher = {Springer}
}

@inproceedings{nisanci_riscv_symm,
  author    = {G{\"o}khan Ni\c{s}anc{\i} and Peter G. Flikkema and Tolga Yal{\c{c}}{\i}n},
  title     = {Symmetric Cryptography on {RISC-V}: Performance Evaluation of Standardized Algorithms},
  booktitle = {Cryptography},
  volume    = {6},
  number    = {3},
  pages     = {41},
  year      = {2022},
  publisher = {MDPI}
}

@inproceedings{stoffelen_riscv,
  author    = {Ko Stoffelen},
  title     = {Efficient Cryptography on the {RISC-V} Architecture},
  booktitle = {Progress in Cryptology -- {LATINCRYPT} 2019},
  series    = {LNCS},
  volume    = {11774},
  pages     = {323--340},
  year      = {2019},
  publisher = {Springer}
}

@inproceedings{adams_riscv_gpgpu,
  author    = {Robert Adams and others},
  title     = {Cryptography Acceleration in a {RISC-V} {GPGPU}},
  booktitle = {CARRV 2021: 5th Workshop on Computer Architecture Research with RISC-V},
  year      = {2021}
}

@inproceedings{rvcp2024,
  author    = {D.~H.~A. Le and others},
  title     = {{RVCP}: High-Efficiency {RISC-V} Co-Processor for Security Applications in {IoT} and Server Systems},
  booktitle = {International SoC Design Conference (ISOCC)},
  year      = {2024},
  publisher = {IEEE}
}

@inproceedings{b2ha_mwscas,
  author    = {Pham Hoai Luan and Thi Sang Duong and Vu Trung Duong Le and Thi Hong Tran and Yasuhiko Nakashima},
  title     = {Energy-Efficient Unified Multi-Hash Coprocessor for Securing {IoT} Systems Integrating Blockchain},
  booktitle = {IEEE 66th Int. Midwest Symp. on Circuits and Systems (MWSCAS)},
  pages     = {355--359},
  year      = {2023}
}

@inproceedings{vrca_icicdt,
  author    = {Vu Trung Duong Le and Hoai Luan Pham and Thi Hong Tran and Quoc Duy Nam Nguyen and others},
  title     = {Versatile Resource-shared Cryptographic Accelerator for Multi-Domain Applications},
  booktitle = {2023 Int. Conf. on {IC} Design and Technology (ICICDT)},
  pages     = {104--107},
  year      = {2023}
}

@inproceedings{rca_mcsoc,
  author    = {Vu Trung Duong Le and Hoai Luan Pham and Thi Hong Tran and Van Duy Tran and Yasuhiko Nakashima},
  title     = {High-efficiency Reconfigurable Crypto Accelerator Utilizing Innovative Resource Sharing and Parallel Processing},
  booktitle = {16th IEEE Int. Symp. on Embedded Multicore/Manycore SoCs (MCSoC)},
  year      = {2023}
}

@inproceedings{1,
author = {Wu, Junwei and Zheng, Xin and Zeng, Shaofen and Gao, Huaien and Xiong, Xiaoming},
title = {High-Performance Cryptographic SoC Virtual Prototyping Platform Based on RISC-V VP},
year = {2022},
isbn = {9781450396295},
publisher = {Association for Computing Machinery},
booktitle = {HP3C '22},
pages = {84–90},
numpages = {7},
location = {Jilin, China},

}

@inproceedings{10.1145/3546000.3546013,
author = {Wu, Junwei and Zheng, Xin and Zeng, Shaofen and Gao, Huaien and Xiong, Xiaoming},
title = {High-Performance Cryptographic SoC Virtual Prototyping Platform Based on RISC-V VP},
year = {2022},
isbn = {9781450396295},
publisher = {Association for Computing Machinery},
booktitle = {HP3C '22},
pages = {84–90},
numpages = {7},
location = {Jilin, China},

}

@ARTICLE{10102649,
  author={Zheng, Xin and Wu, Junwei and Lin, Xian and Gao, Huaien and Cai, Suting and Xiong, Xiaoming},
  journal={TCAS-II}, 
  title={Hardware/Software Co-Design of Cryptographic SoC Based on RISC-V Virtual Prototype}, 
  year={2023},
  volume={70},
  number={9},
  pages={3624-3628},
  doi={10.1109/TCSII.2023.3267186}}

@article{nisanci2022symmetric,
title = {Symmetric Cryptography on RISC-V: Performance Evaluation of Standardized Algorithms},
author = {Ni\c{s}anc\i{}, G"{o}rkem and Flikkema, Paul G. and Yal\c{c}\i{}n, Tolga},
year = {2022},
journal = {Cryptography},
volume = {6},
number = {3},
pages = {41},
}

@InProceedings{10.1007/978-3-030-12942-2_21,
author="Cheng, Hao
and Dinu, Daniel
and Gro{\ss}sch{\"a}dl, Johann",
title="Efficient Implementation of the SHA-512 Hash Function for 8-Bit AVR Microcontrollers",
booktitle="SecITC 2019",
publisher="Springer International Publishing",
address="Cham",
pages="273--287",
isbn="978-3-030-12942-2"
}

@misc{cryptoeprint:2020/1123,
      author = {Alexandre Adomnicai and Thomas Peyrin},
      title = {Fixslicing {AES}-like Ciphers: New bitsliced {AES} speed records on {ARM}-Cortex M and {RISC}-V},
      howpublished = {Cryptology ePrint Archive, Paper 2020/1123},
      year = {2020},
      note = {\url{https://eprint.iacr.org/2020/1123}},
}

@inproceedings{AES2,
author = {Kuo, Yao-Ming and Garcia-Herrero, Francisco and Maestro, Juan Antonio},
year = {2021},
month = {06},
pages = {},
title = {Versatile RISC-V ISA Galois Field arithmetic extension for cryptography and error-correction codes}
}

@INPROCEEDINGS{10908294,
  author={Le, Vu Trung Duong and Tran, Thi Hong Yen and Le, Duc Hong An and Vu, Tuan Hai and Pham, Hoai Luan},
  booktitle={2024 International Conference on Advanced Technologies for Communications (ATC)}, 
  title={RVCP: High-Efficiency RISC-V Co-Processor for Security Applications in IoT and Server Systems}, 
  year={2024},
  volume={},
  number={},
  pages={602-607},
  doi={10.1109/ATC63255.2024.10908294}}

@article{VanTinhNguyen202522.20250329,
  title={AES-RV: Hardware-Efficient RISC-V Accelerator with Low-Latency AES Instruction Extension for IoT Security},
  author={Van Tinh Nguyen and Phuc Hung Pham and Vu Trung Duong Le and Hoai Luan Pham and Tuan Hai Vu and Thi Diem Tran},
  journal={IEICE Electronics Express},
  volume={advpub},
  number={ },
  pages={22.20250329},
  year={2025},
  doi={10.1587/elex.22.20250329}
}
\end{document}